\documentclass[aps,prd,twocolumn,superscriptaddress,floatfix,nofootinbib,natbib]{revtex4-1}
\usepackage{graphicx}
\usepackage{multirow}
\usepackage{latexsym}
\usepackage{enumerate}
\usepackage{footnote}
\usepackage{floatrow}
\usepackage[british]{babel}
\usepackage{soul}
\usepackage{enumitem} 
\usepackage{amsmath}
\usepackage[colorlinks,linkcolor=blue,citecolor=blue,urlcolor=blue ]{hyperref}
\begin{document}

\newcommand{\procspie}{Proceedings of the SPIE}
\newcommand{\aap}{Astronomy and Astrophysics}
\newcommand{\be}{\begin{equation}}
\newcommand{\ee}{\end{equation}}
\newcommand{\bq}{\begin{eqnarray}}
\newcommand{\eq}{\end{eqnarray}}
\newcommand{\bsq}{\begin{subequations}}
\newcommand{\esq}{\end{subequations}}
\newcommand{\bc}{\begin{center}}
\newcommand{\ec}{\end{center}}

\title{Distinguishing between Neutrinos and time-varying Dark Energy through Cosmic Time}

\author{Christiane S. Lorenz }
\email[]{christiane.lorenz@physics.ox.ac.uk}
\affiliation{Sub-department of Astrophysics, University of Oxford, Keble Road, Oxford OX1 3RH, UK}
\author{Erminia Calabrese}
\affiliation{Sub-department of Astrophysics, University of Oxford, Keble Road, Oxford OX1 3RH, UK}
\affiliation{School of Physics and Astronomy, Cardiff University, The Parade, Cardiff, CF24 3AA, UK}
\author{David Alonso}
\affiliation{Sub-department of Astrophysics, University of Oxford, Keble Road, Oxford OX1 3RH, UK}

%\date{\today}

\begin{abstract}
We study the correlations between parameters characterizing neutrino physics and the evolution of dark energy. Using a fluid approach, we show that time-varying dark energy models exhibit degeneracies with the cosmic neutrino background over extended periods of the cosmic history, leading to a degraded estimation of the total mass and number of species of neutrinos. We investigate how to break degeneracies and combine multiple probes across cosmic time to anchor the behaviour of the two components. We use \textit{Planck} CMB data and BAO measurements from the BOSS, SDSS and 6dF surveys to present current limits on the model parameters, and then forecast the future reach from the CMB Stage-4 and DESI experiments. We show that a multi-probe analysis of current data provides only marginal improvement on the determination of the individual parameters and no reduction of the correlations. Future observations will better distinguish the neutrino mass and preserve the current sensitivity to the number of species even in case of a time-varying dark energy component.  

\end{abstract}

\maketitle

%%%%%%%%%%%%%%%%%%%%%%%%%%%%%%%%%%%%%%%%%%%%%%%%%%%%%%%%%%%%%%
\section{Introduction}
Observations from Type Ia Supernovae (SN) \cite{riess1998,Perlmutter1999}, followed by indirect evidence from the Cosmic Microwave Background (CMB) \cite{Sherwin2011,planck2013}, have shown that the expansion of the Universe is accelerating and hint at the existence of an unknown dark energy (DE) component. 

In the standard, concordance cosmological model, dark energy is described in terms of the simplest possible component: a cosmological constant, $\Lambda$, with an equation of state parameter $w_{\rm de}=p_{\rm de}/\rho_{\rm de}$ (pressure over density) constant in time and equal to $-1$. 
However, because of the numerous theoretical issues of the cosmological constant (see e.g., Ref.~\cite{Silvestri:2009hh} and references therein), additional, and more complex, dark energy scenarios have been discussed in the literature, including models in which the dark energy equation of state is varying in time (see e.g., Ref.~\cite{Copeland2006} for a review). While waiting for ongoing and future CMB lensing and galaxy redshift surveys to shed light on the physics of this component, currently available cosmological data are used to constrain all kinds of exotic dark energy models. At present, none of these is a better fit to the data compared to a cosmological constant but also not completely ruled out (see e.g., Ref.~\cite{planck2015xiv} for recent analyses). 

Understanding the nature of dark energy is also particularly relevant for future measurements of parameters characterizing neutrino physics. In particular, Ref.~\cite{allison2015} have performed forecasts for upcoming measurements of neutrino parameters in more extended dark energy scenarios, and have shown that our understanding of neutrinos would be significantly improved if the exact behaviour of dark energy were known. 

Neutrino particles are a key component of the Standard Model of particle physics which accounts for three flavours of very light active particles. The squared mass differences between the neutrino mass eigenstates have been measured by oscillation experiments, $\Delta m_{2,1}^2=7.40\times10^{-5}\text{  eV}$, and $\Delta m_{3,1}^2=2.54\times 10^{-3}\text{eV}$ for the normal hierarchy or $\Delta m_{3,2}^2=2.50\times 10^{-3}\text{  eV}$ for the inverted hierarchy~\cite{patrignani2016}. This leads to a lower limit on the total mass of the three active neutrinos, $\sum m_\nu$, of $59$~meV for the normal hierarchy and $100$~meV for the inverted hierarchy. 
Mass eigenstates limits are also informed by direct neutrino mass searches from laboratory experiments and by cosmological observables: tritium $\beta$-decay experiments set an upper limit on the absolute electron neutrino mass of $2.2$~eV at 95\% confidence (see Ref.~\cite{weinheimer2002} for an overview), and measurements of the growth of cosmic structures from the \textit{Planck} satellite CMB data, combined with baryonic acoustic oscillations (BAO) from low-redshift surveys~\cite{Ross2014,boss2012,wigglez2009}, constrain the total neutrino mass to be $\sum m_\nu \leq 0.21$~eV at 95\% confidence \cite{planck2015xiii}. 

The absolute value of the neutrino mass eigenstates, as well as whether $\Delta m_{3,1}$ and $\Delta m_{3,2}$ are positive or negative and therefore if the neutrino mass hierarchy is normal (positive sign) or inverted (negative sign), are yet to be determined. Improved sensitivity on the absolute electron neutrino mass will soon come from the KATRIN experiment which will reach $m(\nu_e)\sim 0.2$~eV~\cite{drexlin2012}, while future combination of CMB and large-scale structure (LSS) data predict a 4-5$\sigma$ detection of the total neutrino mass with $\sigma(\sum m_\nu) \sim 0.015$~eV in the next decade~\cite{allison2015,abazajian2016,calabrese2016,Louis2016,Core2016}.

In addition to the total neutrino mass, cosmological observations also constrain the effective number of neutrino species, $N_{\rm eff}$, via measurements of the neutrino contribution to radiation density in the early Universe. The current bound on $N_{\rm eff}$ from \textit{Planck} CMB combined with BAO is $3.15\pm0.23$ (at 68\% confidence)~\cite{planck2015xiii}, in agreement with the prediction of the Standard Model of particle physics. An additional constraint on $N_{\rm eff}$ comes from big bang nucleosynthesis (BBN) which limits the number of additional relativistic degrees of freedom at early times to $\Delta N_{\rm rel}\leq1.0$ at 95$\%$ confidence level~\cite{mangano2011}. A 1-2\% determination of $N_{\rm eff}$ is expected from future CMB data~\cite{abazajian2016,Green2016,calabrese2016}. 

At the level of precision of these future measurements, theoretical degeneracies between different cosmological scenarios become important and need to be addressed.
In this paper we investigate in detail the degeneracies between dark energy and neutrino parameters. We show that the main correlations arise if a time-varying dark energy fluid and a neutrino fluid behave very similarly during specific cosmic times and demonstrate that a multi-probe analysis might be able to distinguish between the two. Here, we consider two specific phenomenological dark energy parametrizations (early dark energy and barotropic dark energy) chosen because of their similarity to either the effect of $\sum m_\nu$ or $N_{\rm eff}$, and extend previous analyses presented in Refs.~\cite{calabrese2011a,calabrese2011b}. We use these as a proxy for more general cases and show how to anchor them through cosmic time with a combination of early- and late-time cosmological probes. A multi-probe approach for the specific case of the neutrino mass (without discussing dark energy), and a detailed physical derivation of how to isolate the neutrino mass, has also been presented in Ref.~\cite{Archidiacono2016}. 

The paper is structured as follows. We describe the role of neutrinos in cosmology and the two time-varying dark energy models analysed in this paper in Section \ref{sec:background}. We then present constraints on these models obtained with current CMB and BAO data in Section \ref{sec:methodology}, and forecasts for upcoming experiments in Section \ref{sec:forecasts}. We conclude in Section \ref{sec:conclusion}. 

%%%%%%%%%%%%%%%%%%%%%%%%%%%%%%%%%%%%%%%%%%%%%%%%%%%%%%%%%%%%%%
\section{Theoretical degeneracies}
\label{sec:background}

\begin{figure}[t!]
\includegraphics[width=\columnwidth]{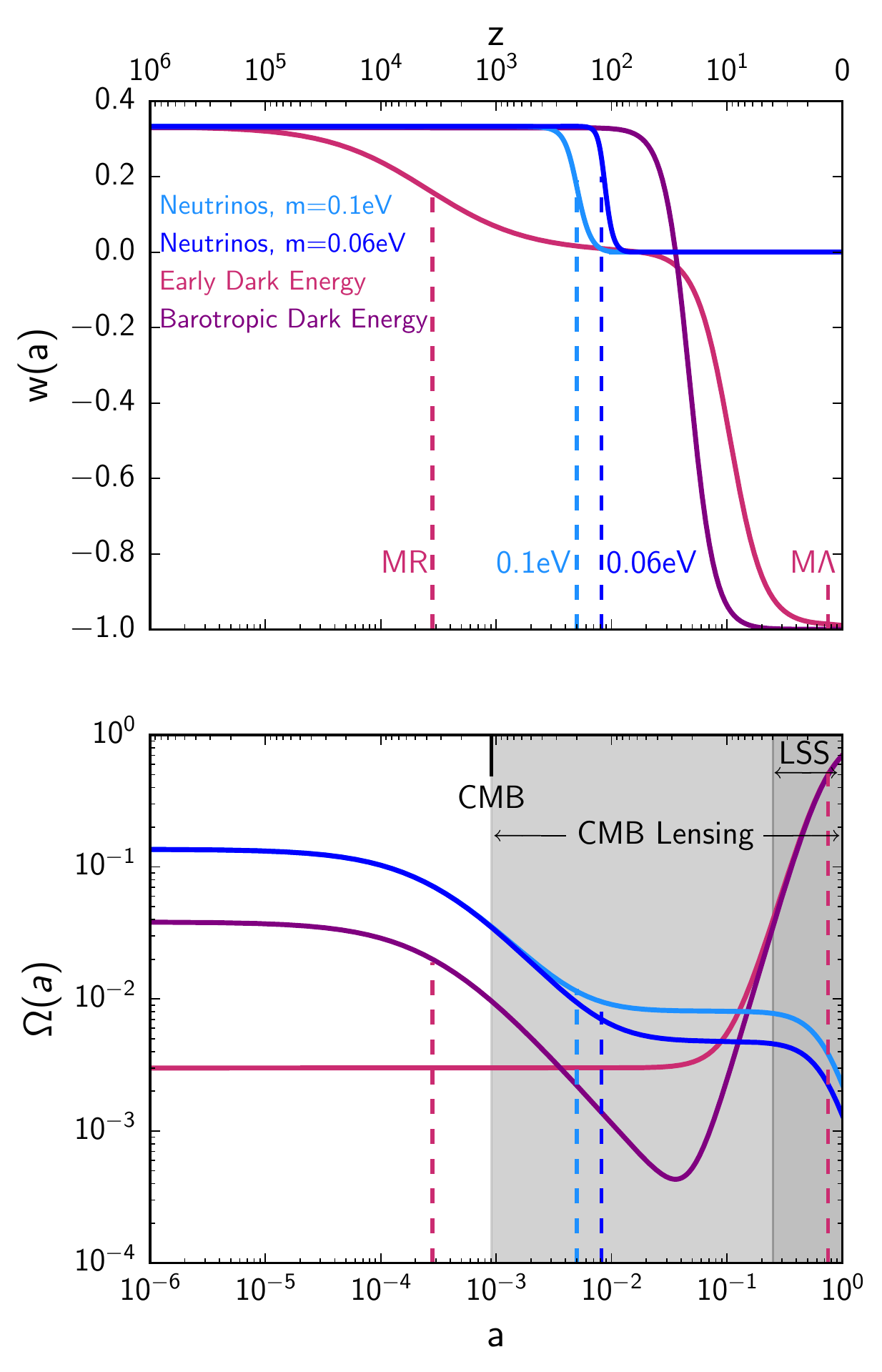}
\caption{\label{fig:comp} The equation of state parameter $w(a)$ (top) and density parameter $\Omega(a)$ (bottom) for neutrinos (blue) and the two specific models of time-varying dark energy~\cite{doran2006,calabrese2011b} (violet) considered in this paper. The x axis in both panels shows the scale factor at the bottom and the corresponding redshift at the top. To generate these predictions we use the standard $\Lambda$CDM \textit{Planck} 2015 best-fit cosmological parameters~\cite{planck2015xiii} in combination with $w_0=-0.99$, $\Omega_e^{\rm EDE}=0.003$, $\Omega_e^{\rm B}=0.038$, $N_{\rm eff}=3.046$ and a single massive neutrino with $\Sigma m_\nu=0.06$ or $0.1$~eV. Dashed vertical lines show the matter-radiation and matter-$\Lambda$ equalities and the time at which a $0.1$~eV and a $0.06$~eV neutrino become non relativistic. The plot also highlights the time at which the CMB decouples and hence which epoch primary CMB anisotropies are probing, and the range probed by CMB lensing and large-scale structure data.}
\end{figure}

Neutrinos and dark energy both affect the expansion rate of the Universe and the growth of cosmic structures, leading to degeneracies between the parameters of the two components even in the case of simple extensions of the cosmological constant (see e.g., Ref.~\cite{hannestad2005,dePutter2009,Font-Ribera2013,Hamann2012,Benoit2012,Pearson:2013iha,Wang:2016tsz,Zhang:2017rbg,Yang:2017amu}). These can be alleviated by combining data which provide orthogonal information in parameter space (an example of this is the measurement of the matter and dark energy densities from galaxy statistics or CMB). Here, we show that a more complicated scenario, with extended degeneracies, arises when dark energy evolves in time with some tracking behaviour. 

To understand phenomenologically why neutrinos and dark energy might look like each other we consider here a fluid parametrization for both components. For each component we define a density parameter, $\Omega(a)\equiv\rho(a)/\rho_{\rm c}(a)$ with $\rho_{\rm c}(a)$ being the critical energy density of the Universe, and an equation of state, $w(a)$, that we evolve with the scale factor, $a$, to track the behaviour of the fluid at different times. We summarize this discussion in Figure~\ref{fig:comp}, which we will gradually populate with models and observational ranges in what follows. 

\subsection{The cosmic neutrino background}
During the history of the Universe neutrinos evolve from a relativistic phase at very early times to a massive-particle behaviour at later times (see Ref.~\cite{lesgourgues2006} for a review). Initially, the neutrinos' kinetic energy dominates over their rest mass energy and as a consequence neutrinos can be considered and described as massless particles fully characterized by their temperature. As the Universe cools down, the kinetic energy decreases and neutrinos transition to a non-relativistic phase with a non-negligible mass. In terms of the energy budget of the Universe, this means that neutrinos contribute to radiation at early times and to matter after the transition, with an energy density given by 
\begin{align}
\rho_\nu(m_\nu\ll T_\nu)&=\frac{7\pi^2}{120}\Big(\frac{4}{11}\Big)^{4/3}N_{\rm eff}T_\gamma^4  \nonumber\\
&= \frac{7}{8}\Big(\frac{4}{11}\Big)^{4/3}N_{\rm eff}\rho_\gamma \,,\nonumber\\
\rho_\nu(m_\nu\gg T_\nu)&=\frac{\rho_{c}}{93.14 h^2 {\rm eV}} \Sigma m_\nu \,
\end{align}
where $T_\nu$ and $T_\gamma$ are the neutrino and photon temperatures, $\rho_\gamma$ is the photon density, and $h$ is the dimensionless Hubble constant. The two parameters of this model are the effective number of relativistic species, $N_{\rm eff}$, and the total mass, $\Sigma m_\nu$.

The transition between the two epochs for the individual neutrino particle happens at a redshift~\cite{ichikawa2004}
\begin{equation}
1+z_{\rm nr} \simeq 120\times\Big(\frac{m_\nu}{60\text{  meV}}\Big) \,.
\end{equation}

In the standard fluid approximation this can be pictured as a time-evolving equation of state $w(a)=p_\nu(a)/\rho_\nu(a)$, which starts from $w_\nu=1/3$ at early times, as for relativistic components, and then subsequently drops to $w_\nu\sim0$ when neutrinos become non relativistic, and as expected for pressure-less matter. The density parameter will reflect this evolution of the individual neutrino particle and manifest distinctive phases as well. This is shown in Figure \ref{fig:comp} with blue lines. \\

\noindent \emph{--The neutrino number--}\\
The Standard Model of particle physics predicts $N_{\rm eff}=3.046$, accounting for the three standard neutrino particles ($\nu_e$, $\nu_\mu$, $\nu_\tau$) and extra energy transfer between neutrinos and the thermal bath as well as QED corrections~\cite{dicus1982,mangano2002,deSalas:2016ztq}. This extra energy is generated during a non-perfectly-instantaneous decoupling of neutrinos from the primordial plasma, with a small part of the entropy released through electron anti-electron annihilations transferred to neutrinos instead of photons. Deviations from the standard predictions will point towards extra radiation in the early Universe or non-standard neutrino decoupling with the initial plasma.

Until the matter-radiation equality, the expansion of the Universe is completely driven by the amount of radiation, which receives contributions from both photons and neutrinos
\begin{equation}
H^2(a)\approx \frac{8\pi G}{3}\Big(\rho_\gamma(a)+\rho_\nu(a)\Big) \,.
\end{equation}

The effective number of neutrinos will then leave an imprint on observables probing $H(a)$ at early times, including the abundances of light elements predicted from BBN, and the CMB primordial temperature and polarization anisotropies. Indeed, the extra energy stored from free-streaming neutrinos at early times delays the time of the matter-radiation equality, and changes the abundances of Helium and Deuterium during BBN. These in turn modify the amplitude, the position and the damping of the CMB anisotropy power spectrum (see, e.g., Refs.~\cite{trotta2003,bashinsky2004,ichikawa2006,Iocco2008,hou2013,archidiacono2011,steigmann2012,follin2015,Green2016,Abazajian2013} for useful discussions). \\

\noindent \emph{--The neutrino mass--}\\
The neutrino mass plays a role only at later times in the history of the Universe. As such, the CMB primordial anisotropies are only mildly affected, but the interaction of the CMB photons with the low-redshift Universe and the large-scale structure formation and growth will have strong signatures of the neutrino mass. 

Since they only interact weakly, neutrinos tend to free-steam out of small-scale density perturbations. As a result, they suppress structure formation on small scales: they do not cluster as a normal matter component would do and they additionally obstacle the cold dark matter and baryon clustering. This can be seen by explicitly comparing the expression of the matter power spectrum, $P(k)$, in the case of massless and massive neutrinos. The power spectrum is suppressed as~\cite{hu1998a}
\begin{equation}
\frac{P(k,\Sigma m_\nu) - P(k, \Sigma m_\nu=0)}{P(k, \Sigma m_\nu=0)}\approx-0.08\Big(\frac{\sum m_\nu}{1 \rm{eV}}\Big)\frac{1}{\Omega_mh^2}
\end{equation}
with $\Omega_m$ being the matter density, for comoving wavelengths larger than $k_{\rm nr}$
\begin{equation}
k_{\rm nr}\approx 0.026\Big(\frac{m_\nu}{1 \rm{eV}}\Big)^{1/2}\Omega_m^{1/2}h\rm{Mpc}^{-1} \,.
\end{equation}

The matter distribution is observationally probed with e.g., measurements of baryon acoustic oscillations, galaxy lensing, and the clustering of the galaxy distribution~\cite{Font-Ribera2013,Abazajian2011}.
The distribution of matter also affects the path of the CMB photons while they travel from the recombination epoch to today: gravitational potential wells along the photons' path will generate small deflections in the CMB temperature and polarization anisotropies and produce a CMB weak-lensing signal~\cite{lewis2006}. CMB lensing will therefore reflect the matter power spectrum dependence on the neutrino mass (with massive neutrinos suppressing the overall amplitude of the CMB lensing signal) and will be an indirect probe for it (see e.g., Refs.~\cite{dePutter2009,planck2013,planck2015xiii,allison2015,Sherwin2016}). 

\subsection{Time-varying dark energy}
To study the evolution of the Universe in the presence of more complicated dark energy models, we implemented two phenomenological parametrizations described below. The choice of the models is based on their interesting, and at the same time problematic, similarity to the neutrino fluid evolution. For both models we included a full set of perturbation equations with constant sound speed and viscosity parameters equal to $1/3$. This choice of parameters is made to highlight the degeneracies with the neutrino sector and is discussed in detail in Ref.~\cite{calabrese2011b}. 

\subsubsection{Barotropic dark energy}
The barotropic class of dark energy models~\cite{linder2009} include all sorts of models in which the physics of the dark energy fluid is fully determined by the pressure as an explicit function of the density. The key feature of these models is the simple extension of the cosmological constant to a theory where DE is varying in time through a non-zero DE term present at early times and then quickly transitioning to $\Lambda$ today. This alleviates the $\Lambda$ fine-tuning problem and is still in agreement with current cosmological data.

One such example is the model presented in Ref.~\cite{linder2009} where the DE equation of state is given by

\begin{equation}
\label{eq:wbar}
w_{\rm baro} (a)=[c_s^2Ba^{-3(1+c_s^2)}-1]/[Ba^{-3(1+c_s^2)}+1] \,,
\end{equation}
where $c_s$ is the dark energy sound speed, $B=(1+w_0)/(c_s^2-w_0)$, and $w_0$ is the present value of the equation of state. We extend this model by introducing perturbations in the DE fluid as in Ref.~\cite{calabrese2011b} and in fact continuing the late-time DE term with a dark radiation term at early times. To discuss the interesting degeneracies with neutrino physics we fix $c_s^2=1/3$. Consequently $w$ goes to $1/3$ for $a\rightarrow 0$, as in the case of radiation and neutrinos, and approaches $w=-1$ today. 

The barotropic dark energy density is now obtained by inserting Eq.~\ref{eq:wbar} in the dark energy continuity equation and integrating this latter to obtain
\begin{equation}
\rho_{\rm baro}(a)=\frac{\rho_{\rm baro,0}}{B+1}(1+Ba^{-4}) \,,
\end{equation}
where the subscript 0 stands for today and with $\rho_{\rm baro,0}$ given by 
\begin{equation}
\rho_{\rm baro,0}=\Big(\frac{3H_0^2}{8\pi G}\Big)\times(1-\Omega_{\rm m,0}) \,.
\end{equation} 

In this model the dark energy fluid can be approximated as the sum of a late-time cosmological constant and an additional radiation term dominating at early times, $\rho_{\rm baro} \sim \rho_\infty + A a^{-4}$. 

The fraction of barotropic dark energy contributing to radiation in the early Universe depends on the only free parameter of the model, $B$, and can be computed through
\begin{equation}
\begin{split}
\Omega_e^{\rm B}&=\lim_{a\rightarrow 0}\frac{\rho_{\rm baro}(a)}{\rho_{\rm c}(a)}\\
&=\frac{B\rho_{\rm baro,0}}{B\rho_{\rm baro,0}+(B+1)\rho_{\rm r,0}}
\end{split}
\end{equation}
where $\rho_{\rm r,0}$ is the radiation density today (assuming $N_{\rm eff}$ massless neutrinos)\footnote{We note that we have derived here a different parametrization of $\rho_{\rm baro}(a)$ and therefore a new derivation of $\Omega_e^{\rm B}$, which does not directly correspond to the same parameter in Ref.~\cite{calabrese2011b}.}.

The density and equation of state for this model in the case of $B=5\times10^{-6}$ ($\Omega_e^{\rm B}\sim 0.038$) are shown in Figure~\ref{fig:comp} in dark violet lines. By construction, this model is now degenerate with the neutrino fluid during radiation domination and, because of this, we expect correlations between $B$ (or equivalently $\Omega_e^{\rm B}$) and $N_{\rm eff}$.

\subsubsection{Early dark energy}
The second model that we consider is an early dark energy model (EDE) that has been first suggested by Ref.~\cite{doran2006} and extensively explored in the literature~\cite{dePutter2010,calabrese2011a,calabrese2011b,Reichardt2012,Pettorino2013,Calabrese2014,planck2015xiv}. This model falls into the tracking dark energy class of models ~\cite{Ferreira&Joyce1997}, where the dark energy density is a sub-dominant fraction of the dominant component of each cosmic epoch, i.e., radiation first, matter later and evolving into $\Lambda$ today. 

The dark energy density and the equation of state parameters are given by

\begin{align}
\Omega_{\rm ede}(a)&=\frac{\Omega_{\rm ede,0}-\Omega_{e}^{\rm EDE}(1-a^{-3w_0})}{\Omega_{\rm ede,0}+\Omega_{\rm m,0}a^{3w_0}}+\Omega_{\rm e}^{\rm EDE}(1-a^{-3w_0}) \,, \\
w_{\rm ede}(a)&= -\frac{1}{3[1-\Omega_{\rm ede}(a)]}\frac{d\ln{\Omega_{\rm ede}(a)}}{d\ln{a}}
+\frac{a_{\rm eq}}{3(a+a_{\rm eq})} \,,
\end{align}
and shown in Figure~\ref{fig:comp}. The two free parameters of the model are the present value of the equation of state parameter, $w_0$, and $\Omega_{\rm e}^{\rm EDE}$, which is the asymptotic limit of the DE energy density at $a=0$. $a_{\rm eq}$ is the scale factor at matter-radiation equality. The evolution of $w$ in this case is more complex and can be divided into three regimes: $w\simeq1/3$ during radiation domination, $w\simeq0$ during matter radiation and $w=w_0$ today. 

In this case, as is clear from Figure~\ref{fig:comp}, $w_{\rm ede}$ transitions to a matter-like behaviour before neutrinos become non relativistic and the two fluids are degenerate at early-to-intermediate times. Hence, the early dark energy model parameters will be mostly correlated with the neutrino mass. In particular, in the late Universe, both early dark energy and neutrinos now suppress structure formation: neutrinos through the effect of their mass described before, and dark energy by changing the expansion rate~\cite{dePutter2009}.

\subsection{Observations at different cosmic times}
The above discussion and Figure~\ref{fig:comp} stress the need to test cosmological models at different cosmic epochs to distinguish between neutrinos and time-varying dark energy, with observations spanning a wide range of redshifts. This is possible combining measurements of the early Universe via the CMB primary anisotropies with large-scale structure data measuring the late-time evolution (including galaxy weak lensing and clustering, baryonic acoustic oscillations and SN distance measurements), connected at intermediate times via the CMB gravitational lensing. This is schematically shown in Figure~\ref{fig:comp}, where we highlight the time of the CMB decoupling (redshift of $z\simeq 1100$), and where CMB lensing (integrated signal from decoupling to today) and LSS ($3\gtrsim z>0$) sit relative to the evolution of massive neutrinos and a time-varying dark energy. The blue dashed lines show the time when a $0.1$~eV and a $0.06$~eV neutrino become non-relativistic, at $z\simeq 5\times10^{-3}$ and $z\simeq 8\times 10^{-3}$, respectively. Magenta dashed lines show the times of the matter-radiation and the matter-$\Lambda$ equality defining the DE transitions. 

In particular, for the models considered here, the pattern of acoustic peaks in the CMB primary power spectra will anchor the relativistic behaviour and so provide information on $N_{\rm eff}$ and $\Omega_e^{\rm B}$, while CMB lensing and LSS will distinguish the fluids in the matter- and $\Lambda$-dominated epochs, improving the limits on $N_{\rm eff}$ and $\Omega_e^{\rm B}$, and constraining $\sum m_\nu$, $w_0$ and $\Omega_e^{\rm EDE}$.  

This multi-probe combination has already proven to be very powerful in testing cosmological models~\cite{planck2015xiii} and will become a standard approach for future analyses of CMB and LSS data. Anticipating high-precision and high-sensitivity CMB primary and lensing observations from the ground-based CMB Stage IV experiment~\cite{abazajian2016}, and their combination with BAO from the Dark Energy Spectroscopic Instrument (DESI)~\cite{desi2013}, or galaxy lensing and clustering from the Large Synoptic Survey Telescope (LSST)~\cite{lsst}, the Euclid satellite~\cite{euclid} and the Wide-Field InfraRed Survey Telescope (WFIRST) mission~\cite{wfirst}, we investigate in the following current limits and future prospects for these models.

%%%%%%%%%%%%%%%%%%%%%%%%%%%%%%%%%%%%%%%%%%%%%%%%%%%%%%%%%%%%%%
\section{Constraints from current data}
\label{sec:methodology}
To constrain the dark energy model parameters in conjunction with neutrino physics with current CMB and LSS data, we modified a publicly available version of the CAMB Boltzmann code~\cite{lewis2000} and interfaced it with CosmoMC \cite{lewis2002}, a public Monte Carlo Markov chain package that explores cosmological parameters for different theoretical models and data combinations. 

We explore an extended $\Lambda$CDM model where we vary the standard cosmological parameters (the baryon density today $\Omega_{\rm b}h^2$, the cold dark matter density today $\Omega_{\rm c}h^2$, the scalar spectral index $n_s$, the Hubble constant $H_0$, the amplitude of primordial scalar perturbations $A_s$) and additional DE and neutrino parameters: $N_{\rm eff}$, $\Sigma m_\nu$, $\Omega_e^{\rm EDE}$, $w_0$ (in the range $w_0>-1$), and $B$. For this latter parameter we impose a flat prior in the range [-7,-2] on its logarithmic variation to better explore very small values, and we report results in terms of its derived parameter $\Omega_e^{\rm B}$. When not varied, we follow the standard convention of fixing $N_{\rm eff}=3.046$, $\Sigma m_\nu=0.06$~eV, $\Omega_e^{\rm EDE}=0$, $w_0=-1$, and $B=0$.
We further impose a Gaussian prior on the reionization optical depth, $\tau=0.06\pm0.01$, in order to incorporate recent CMB large-scale polarization data from the \textit{Planck} satellite~\cite{planck2016xlvi}. 

We extract cosmological parameters using CMB primary and lensing data from the \textit{Planck} 2015 data release~\cite{planck2015xi,planck2015xv} (retaining only high-multipole temperature for primary anisotropies as recommended by the \textit{Planck} team), and BAO distance ratio $r_s/D_V$ from BOSS DR12 (CMASS and LOWZ)~\cite{boss2015}, SDSS MGS~\cite{Ross2014}, and 6DF~\cite{6df2011}. We further impose the BBN consistency relation between $N_{\rm eff}$ and the baryon density on the primordial Helium abundance~\cite{Pisanti2007}. 

\subsubsection{Single-probe degeneracies}

\begin{figure}[t!]
\includegraphics[width=\columnwidth]{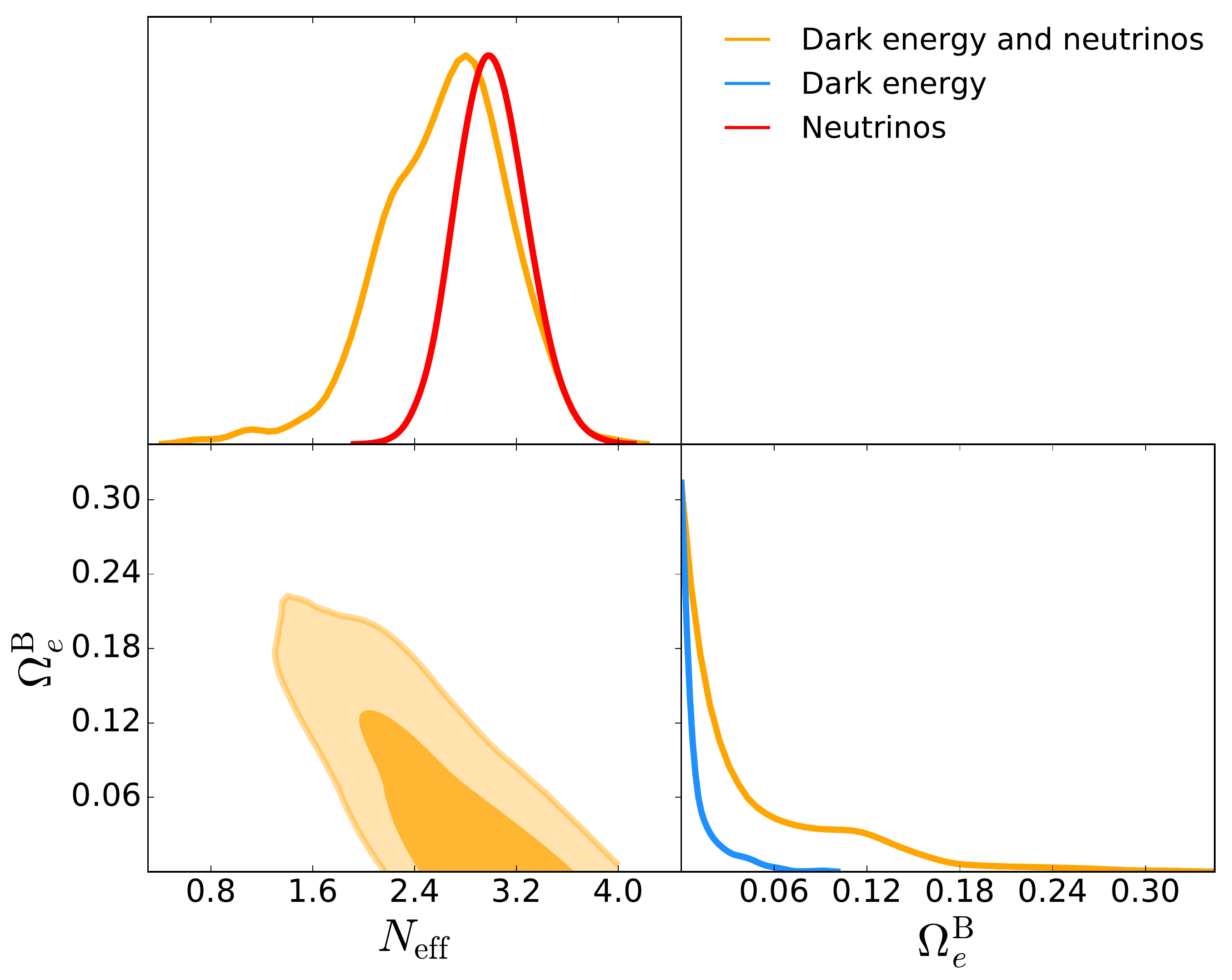}
\vspace*{-3mm}
\caption{1-dimensional posterior and 2-dimensional contour levels at 68\% and 95\% confidence for the effective number of neutrinos, $N_{\rm eff}$, and the early barotropic dark energy density, $\Omega_e^{\rm B}$, constrained by \textit{Planck} CMB temperature anisotropies. Different colours distinguish runs with different freedom in the parameters space: red for varying only the neutrino parameters, blue for varying only dark energy ones, and orange for parameters of both components varying at the same time.}\label{fig:bar_planck}
\end{figure}

\begin{figure}[t!]
\includegraphics[width=\columnwidth]{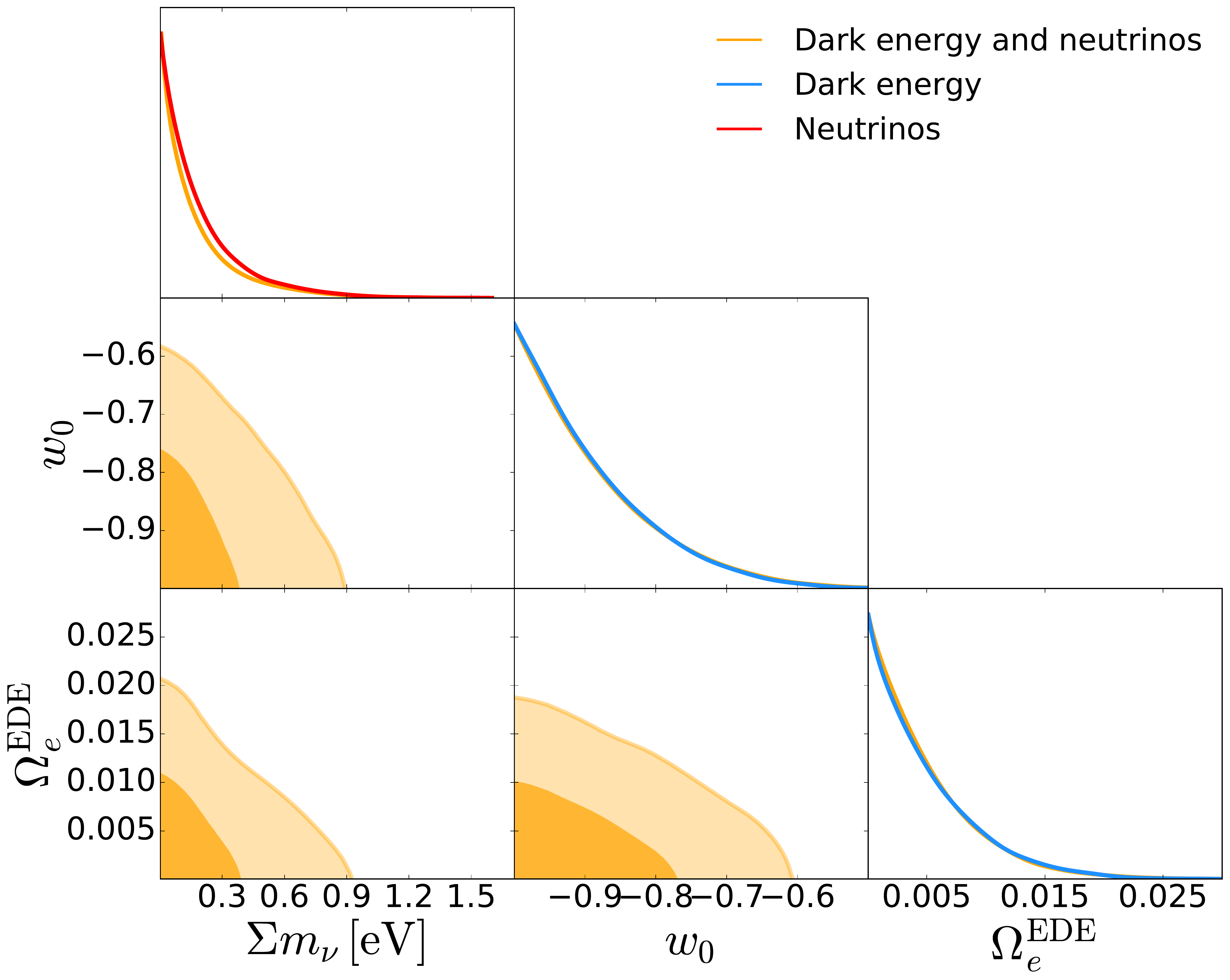}
\vspace*{-3mm}
\caption{1-dimensional posterior and 2-dimensional contour levels at 68\% and 95\% confidence in the case of degeneracies between massive neutrinos and the early dark energy model. The parameters varied are the neutrino mass sum, $\sum m_\nu$, the early dark energy density, $\Omega_e^{\rm EDE}$, and the present value of the DE equation of state, $w_0$. The colour scheme is the same as Figure~\ref{fig:bar_planck}.}\label{fig:ede_planck}
\end{figure}

We first consider the case in which a single probe is used to constrain time-varying DE and neutrinos. For this we retain the most constraining probe of the Universe's content and evolution, the primary CMB anisotropies. Limits from \textit{Planck} CMB temperature data are shown in Figures~\ref{fig:bar_planck},~\ref{fig:ede_planck}, where we recover the expected $N_{\rm eff}-\Omega_e^{\rm B}$, $\sum m_\nu - \Omega_e^{\rm EDE} - w_0$ degeneracies.


To show the impact of one component on the other, we run three different cases for each of the time-varying DE models: (i) opening only neutrino parameters, (ii) opening only DE parameters, (iii) varying all DE and neutrino parameters at the same time. We report quantitative results in terms of the correlation coefficient, defined as
\begin{equation}
R = C(P_1,P_2)/\sqrt{(C(P_1, P_1)\times C(P_2, P_2))} \,,
\end{equation}
where C is the covariance matrix of the P parameters. 

In the case of (i) we recover the \textit{Planck} limits on $N_{\rm eff}$ and $\Sigma m_\nu$~\cite{planck2016xlvi}, yielding $N_{\rm eff}=3.00\pm 0.28$ (68\% confidence) and $\sum  m_\nu< 0.63$~eV (at 95\% confidence). 

The individual DE parameters in the case of (ii) are instead constrained to be: $\Omega_e^{\rm B}<0.045$, $\Omega_e^{\rm EDE}<0.014$, and $w_0<-0.72$ (all at 95\% confidence), where the latter two are consistent with the \textit{Planck} results in Ref.~\cite{planck2015xiv}.

When letting both components free to vary we see that the limits on the individual parameter degrade by $77\%$ for $N_{\rm eff}$ and $284\%$ for $\Omega_e^{\rm B}$, and a correlation of $-81\%$ is found between the two. To fit the \textit{Planck} high-precision CMB acoustic peaks position, the amount of radiation is split between $N_{\rm eff}$ and $\Omega^{\rm B}_e$ along a tightly constrained anti-correlated region. 

The impact on individual constraints is instead less strong in the case of $\Omega_e^{\rm EDE} - w_0 - \sum m_\nu$. This can be understood by noticing that in this case the results are dominated by the sampling and physical priors ($\Omega_e^{\rm EDE}>0$, $w_0>-1$, and $\sum m_\nu>0$) which confines all the parameters into the lower limit region of the samples and hides the anti-correlation (see Figure~\ref{fig:ede_planck}). We will show that this will not be the case with future data, when one of the parameters (the neutrino mass sum in this case) will be constrained away from the sampling bounds.

\subsubsection{Multi-probe analysis}

\begin{table}[t]
\begin{tabular}{|l|c|c|c|}
\hline 
Parameters & CMB & CMB & CMB \\
 & & +CMBL & +CMBL+BAO \\
\hline
\emph{Baro DE} & & & \\
$\Omega_e^{\rm B}$ & $\leq0.164$ & $\leq0.115$ & $\leq0.107$\\
$N_{\rm eff}$ & $2.64\pm0.49$ & $2.83\pm0.35$ & $2.87\pm0.33$ \\
\hline
\emph{EDE} & & & \\
$\Omega_e^{\rm EDE}$ & $\leq0.013$ & $\leq0.011$ &  $\leq0.007$\\
$w_0$ & $\leq-0.71$ & $\leq -0.73$ & $\leq-0.89$ \\
$\sum m_\nu$[eV] & $\leq0.64$ & $\leq 0.53$ & $\leq0.12$ \\
\hline 
\hline
Correlations & CMB & CMB & CMB \\
 & & +CMBL & +CMBL+BAO \\
\hline
\emph{Baro DE} & & & \\
$N_{\rm eff}-\Omega_e^{\rm B}$ & -81\% & -66\% & -79\%\\
\hline
\emph{EDE} & & & \\
$\sum m_\nu-\Omega_e^{\rm EDE}$ & -3.7\% & -20\% &-15\%\\
$\sum m_\nu-w_0$ & 2.3\% & 0.3\% & -19\%\\
\hline
\end{tabular} 
\caption{\label{tab:ede} \emph{Top:} Marginalized constraints on dark energy and neutrino parameters for different data combinations: \textit{Planck} primary CMB and CMB lensing (CMBL), and BAO data probing the large-scale structure. Errors are 68\% confidence levels while upper limits are reported at 95\% confidence. \emph{Bottom:} Correlation coefficients between the DE and neutrino model parameters for different data combinations.}
\end{table}

\begin{figure}[t!]
\includegraphics[width=\columnwidth]{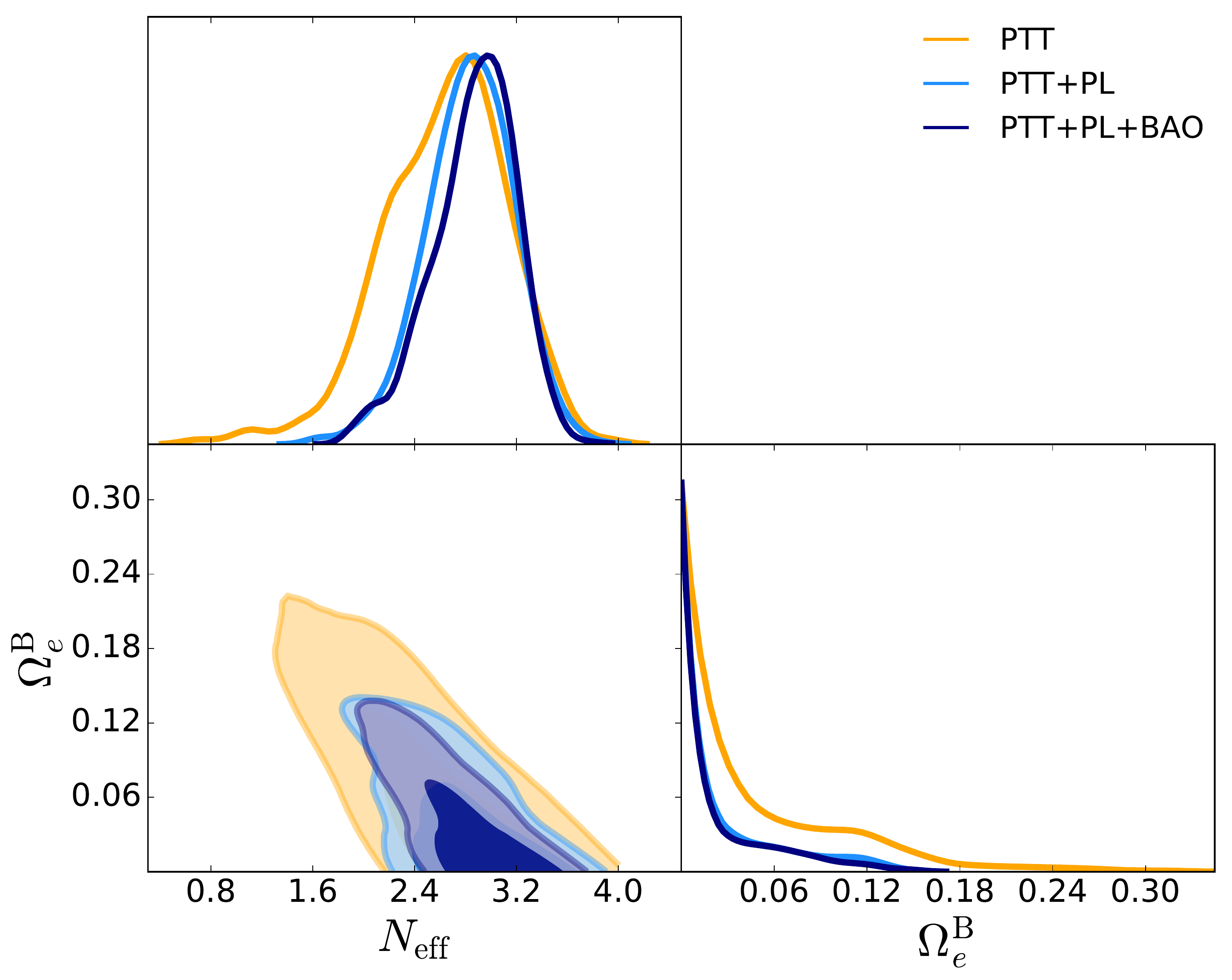}
\vspace*{-3mm}
\caption{1-dimensional posterior and 2-dimensional contour levels at 68\% and 95\% confidence for the effective number of neutrinos, $N_{\rm eff}$, and the early barotropic dark energy density, $\Omega_e^{\rm B}$, from different data combinations: \textit{Planck} CMB primary anisotropies (PTT) only in orange, combined with \textit{Planck} CMB lensing spectra (PL) in light blue, and with also BAO in dark blue. The inclusion of low-redshift data helps only marginally to reduce the degeneracies between $N_{\rm eff}$ and $\Omega_e^{\rm B}$.}\label{fig:bar}
\end{figure}

\begin{figure}[t!]
\includegraphics[width=\columnwidth]{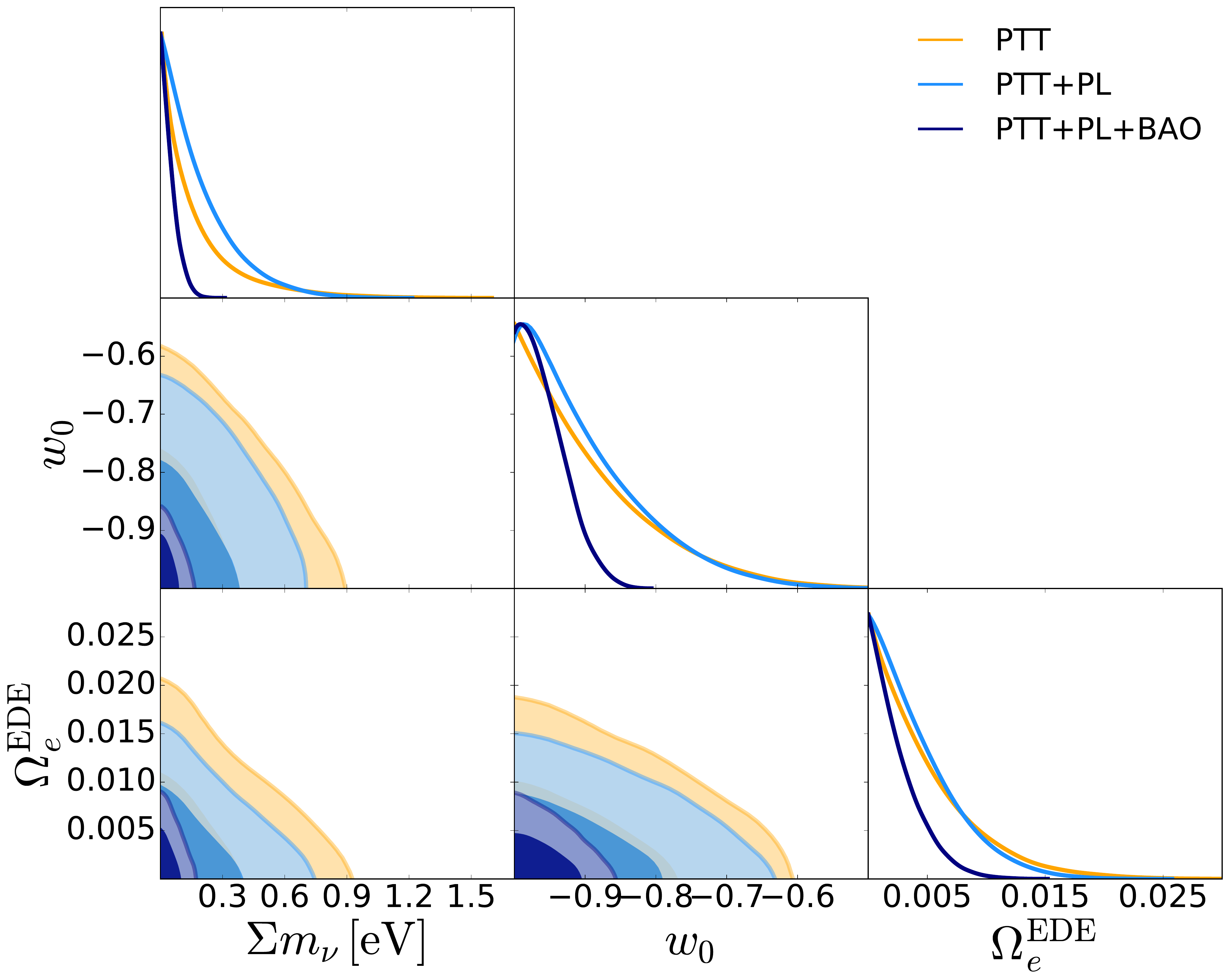}
\vspace*{-3mm}
\caption{1-dimensional posterior and 2-dimensional contour levels at 68\% and 95\% confidence for the sum of the neutrino masses, $\Sigma m_\nu$, the early dark energy density, $\Omega_e^{\rm EDE}$, and the present value of the dark energy equation of state, $w_0$, from different data combinations: \textit{Planck} CMB primary anisotropies (PTT) only in orange, combined with \textit{Planck} CMB lensing spectra (PL) in light blue, and with also BAO in dark blue. The inclusion of low-redshift data helps to reduce the degeneracies between $\sum m_\nu$ and $\Omega_e^{\rm EDE}-w_0$.}\label{fig:ede}
\end{figure}

To show how a multi-probe analysis can help confine the two components and hence break the degeneracies, we report the results of gradually adding to the main \textit{Planck} CMB primary spectra late-time probes, including \textit{Planck} CMB lensing, and BOSS/SDSS/6dF BAO. State-of-the-art constraints on these models are reported in Table~\ref{tab:ede} and Figures~\ref{fig:bar},~\ref{fig:ede}. 

In the case of the barotropic dark energy model, low-redshift data only marginally improve individual parameters constraints and do not help in reducing the correlations. This can be understood considering that both $B$ and $N_{\rm eff}$ are mainly constrained via the expansion rate at very early times. Primary CMB is then dominating the constraints, with CMB lensing providing some additional contribution at intermediate redshifts and no extra information coming from BAO. 

For the second scenario (early dark energy), low-redshift data have a stronger impact by providing tight bounds on the matter component. Because of this, the sum of the neutrino masses and the amount of $\Omega_e^{\rm EDE}$ are better constrained. They also provide a much tighter constraint on $w_0$, helping to better limit the 3-dimensional degeneracy $\sum m_\nu - \Omega_e^{\rm EDE} - w_0$. We have also tested whether the inclusion of the Type Ia Supernovae compilation of the Joint Light-curve analysis (JLA) team ~\cite{jla2013} helps to better constrain the early dark energy model, but found no significant improvement.

Table~\ref{tab:ede} also reports the values of the correlation coefficient for both scenarios and all data combinations. While the combination of CMB and BAO (as a LSS probe) improves the individual parameters' constraints, the current level of sensitivity is not able to isolate and then break the correlations in two dimensions.

%%%%%%%%%%%%%%%%%%%%%%%%%%%%%%%%%%%%%%%%%%%%%%%%%%%%%%%%%%%%%%
\section{Future predictions}
\label{sec:forecasts}

To estimate the power of future cosmological data in distinguishing between neutrinos and these time-varying dark energy models, we present here predictions of future limits using the CMB Stage-4 experiment (S4) in combination with BAO measurements from DESI as a tracer of the large-scale structure\footnote{We note that a different LSS tracer would lead to the same qualitative conclusions.}. 
From mid-2020s we anticipate access to arcminute-resolution CMB temperature and polarization data with a $1\mu$K-arcmin noise level from CMB-S4~\cite{abazajian2016}, and percent-level determination of the Hubble constant and angular diameter distance from DESI~\cite{Font-Ribera2013}, tracing the history of the Universe with unprecedented sensitivity.

We run Fisher matrix analyses using the code presented in Ref.~\cite{Alonso2015} and following the methodology described in Ref.~\cite{calabrese2016} for the data combination (see Table I in there). Our reference datasets are: 
\begin {itemize}
\item[-]{PL+S4}\\
CMB-S4 temperature and E-modes of polarization anisotropies over $30<\ell<3000/5000$ on 40\% of the sky measured with a $3$-arcmin resolution and $1\mu$K-arcmin noise level in temperature; combined with expected full-mission \textit{Planck} data (as implemented in Refs.~\cite{allison2015,calabrese2016}) to complement the multipole range and extend the sky fraction;
\item[-]{PL+S4+S4L}\\
same as above but including also CMB-S4 measurements of the CMB lensing power spectrum over $30<\ell<3000$ on 40\% of the sky;
\item[-]{PL+S4+S4L+BAO}\\
same as above plus BAO distance ratio as measured by DESI in the range $0.15<z<1.85$.
\end{itemize}

\begin{figure}[t!]
\includegraphics[width=\columnwidth]{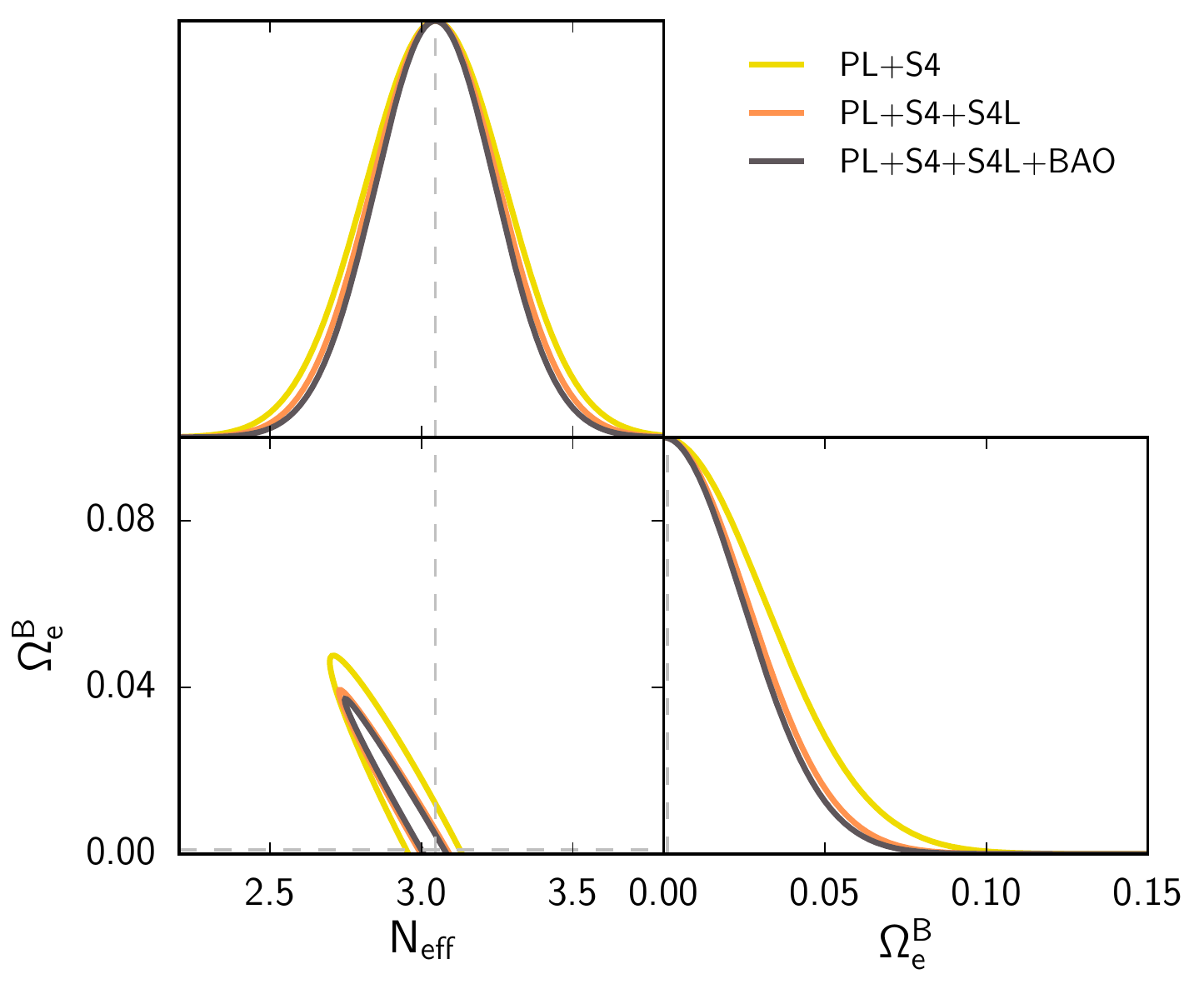}
\vspace*{-3mm}
\caption{Predictions for the constraints on the barotropic dark energy density, $\Omega_e^{\rm B}$, and the effective number of neutrinos, $N_{\rm eff}$,  from future CMB-S4 primary anisotropies (S4) and lensing (S4L) data, complemented by \textit{Planck} (PL), and in combination with BAO distance ratio from DESI. The 2-dimensional contours report the 68\% confidence levels. The dashed lines show the fiducial values of the parameters used in the Fisher calculations.}\label{for_bar}
\end{figure}

\begin{figure}[t!]
\includegraphics[width=\columnwidth]{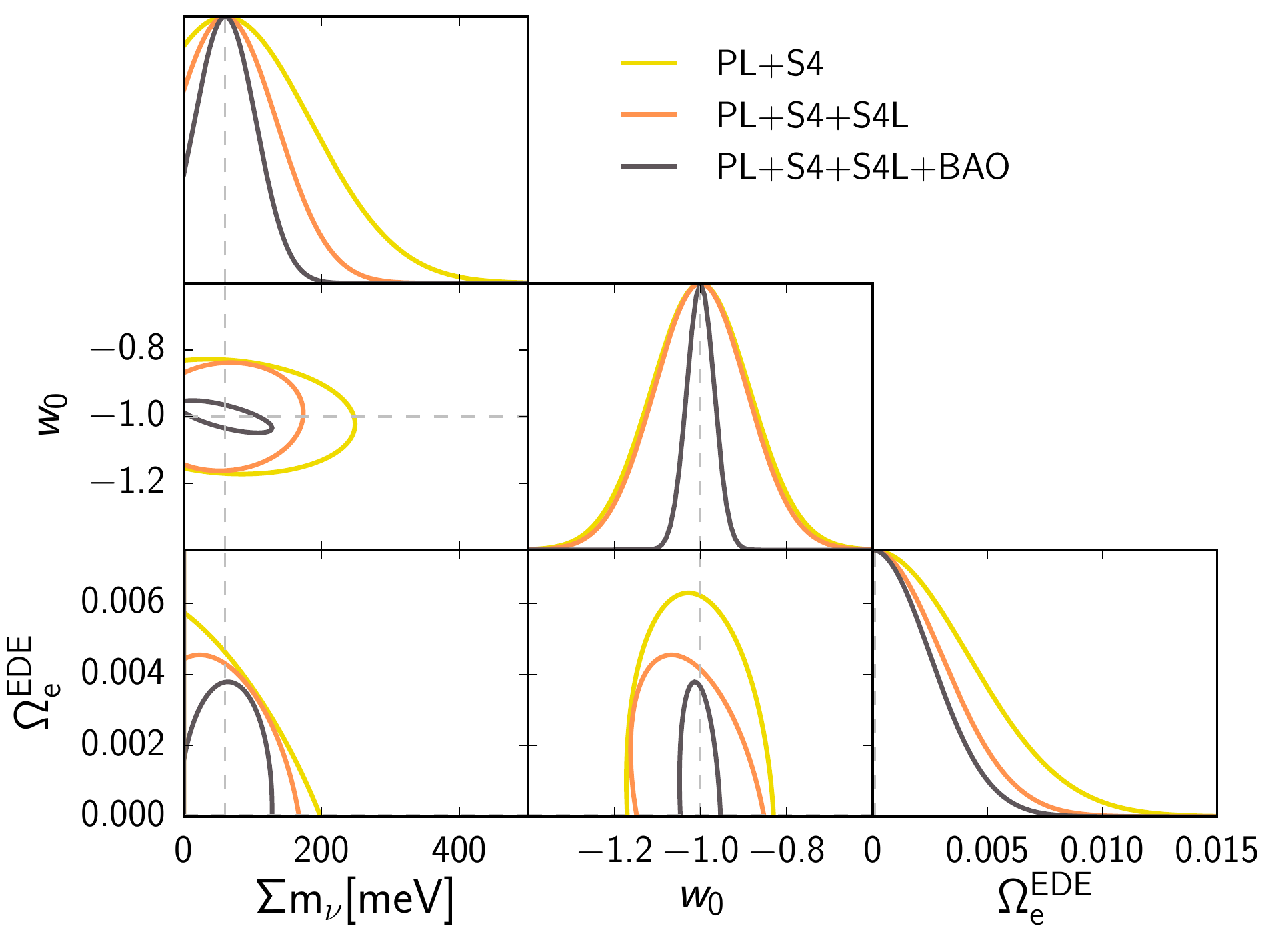}
\vspace*{-3mm}
\caption{Same as Figure~\ref{for_bar} in the case of early dark energy; showing future constraints on the dark energy density, $\Omega_e^{\rm EDE}$, the present value of the equation of state parameter, $w_0$, and the neutrino mass sum, $\Sigma m_\nu$, obtained from different early- and late-time data combinations.}\label{for_ede}
\end{figure}

The results are shown in Figures~\ref{for_bar},~\ref{for_ede} for barotropic and early dark energy, respectively. 


In the case of barotropic dark energy, future CMB data will significantly improve the constraints on the individual parameters, reaching the current level of sensitivity for $N_{\rm eff}$ in the case of no varying dark energy ($\sigma(N_{\rm eff})\sim0.2$ from \textit{Planck}+BAO) and limiting the fraction of barotropic dark energy at early times with percent-level accuracy. The high correlation between $N_{\rm eff}$ and $\Omega_e^{\rm B}$, however, persists even with higher-resolution data
\begin{eqnarray}
R(N_{\rm eff},\Omega_e^{\rm B}) & = & -97\% \quad \text{(PL+S4)} \,, \nonumber\\
& = & -97\% \quad \text{(PL+S4+S4L)} \,, \nonumber\\
& = & -99\% \quad \text{(PL+S4+S4L+BAO)} \,.
\end{eqnarray}

Ref.~\cite{Green2016} have shown a $\sim20$\% improvement on the determination of $N_{\rm eff}$ when BBN information are added to CMB-S4 by, e.g., imposing BBN consistency relations. We choose not to include BBN information here because it would not change our conclusions. In the presence of barotropic dark energy, the addition of BBN would be less effective in constraining $N_{\rm eff}$ and not useful to break the degeneracies with DE. Barotropic DE would in fact affect the BBN just as extra relativistic degrees of freedom (an effective $\Delta N_{\rm eff}$) and therefore will continue to mimic neutrino particles all the way to the BBN epoch. We note that this is due to our way of defining the two fluids with the same sound speed $c_s^2=1/3$ and viscosity parameter $c_{\rm vis}^2=1/3$, which therefore cannot be isolated with higher-order velocity/viscosity propagation. In the case of non free-streaming extra radiation, a measurement of the phase shift in the CMB anisotropies will break these correlations (see, e.g., Refs.~\cite{bashinsky2004,follin2015,Baumann2016}).

The multi-probe approach is instead very successful for the early dark energy scenario. Figure~\ref{for_ede} shows a decreasing correlation between the parameters (visually appreciated in the rotation of the 2-dimensional contours) with the addition of lower-redshift data. The correlation coefficient is found to be 

\begin{eqnarray}
R(\Sigma m_\nu,\Omega_e^{\rm EDE}) & = & -68\% \quad \text{(PL+S4)} \,, \nonumber\\
& = & -33\% \quad \text{(PL+S4+S4L)} \,, \nonumber\\
& = & +6.0\% \quad \text{(PL+S4+S4L+BAO)} \,,
\end{eqnarray}

\begin{eqnarray}
R(\Sigma m_\nu,w_0) & = & -13\% \quad \text{(PL+S4)} \,, \nonumber\\
& = & -7.0\% \quad \text{(PL+S4+S4L)} \,, \nonumber\\
& = & -7.5\% \quad \text{(PL+S4+S4L+BAO)} \,. 
\end{eqnarray}

In the presence of time-varying dark energy, the estimate of the neutrino mass is therefore significantly aided by combining multi-epoch datasets. We find for PL+S4+S4L+BAO $\sigma(\Sigma m_\nu)\sim 0.04$~eV, which is a $\sim1.5$ factor worse than CMB-S4 predictions in a $\Lambda$CDM scenario when combined with DESI. This will improve even more when Supernovae, galaxy shear and clustering, galaxy cluster counts and redshift space distortions are optimally combined with the probes we considered here.

%%%%%%%%%%%%%%%%%%%%%%%%%%%%%%%%%%%%%%%%%%%%%%%%%%%%%%%%%%%%%%

\section{Conclusion}
\label{sec:conclusion}
In this paper we have investigated the correlations arising between time-varying dark energy models and cosmological neutrinos. We have demonstrated how some dark energy models tracking other cosmic components during specific epochs can look like neutrinos over extended periods of the Universe history. This will affect our ability to constrain the number and sum of the masses of the neutrino particles and the physics of dark energy. 

We have considered two phenomenological dark energy models: barotropic dark energy and early dark energy, particularly interesting due to their similarity to the effects on cosmological probes of either $N_{\rm eff}$ or $\Sigma m_\nu$. We have presented state-of-the-art limits on these models but found that current CMB and large-scale structure data are not able to clearly distinguish between the two components. In addition, we have investigated the reach of future experiments and forecast estimates from the CMB Stage-4 experiment in combination with BAO from DESI. We have shown that future data will be able, via a multi-probe combination, to break some of the degeneracies and better limit these extended scenarios.

%%%%%%%%%%%%%%%%%%%%%%%%%%%%%%%%%%%%%%%%%%%%%%%%%%%%%%%%%%%%%%
\begin{acknowledgments}
We thank Eric Linder and Dan Green for useful discussions. CL is supported by a Clarendon Scholarship and acknowledges support from Pembroke College, Oxford. EC is supported by a Science and Technology Facilities Council (STFC) Rutherford Fellowship. DA is supported by the STFC and the Beecroft Trust.
\end{acknowledgments}
%%%%%%%%%%%%%%%%%%%%%%%%%%%%%%%%%%%%%%%%%%%%%%%%%%%%%%%%%%%%%%

%%%%%%%%%%%%%%%%%%%%%%%%%%%%%%%%%%%%%%%%%%%%%%%%%%%%%%%%%%%%%%
\bibliography{ref}

\end{document}